# Identifying the electronic character and role of the Mn states in the valence band of (Ga,Mn)As

J. Fujii[1], B. R. Salles[1], M. Sperl[2], S. Ueda[3], M. Kobata[3], K. Kobayashi[3,4,5], Y. Yamashita[3,6], P. Torelli[1], M. Utz[2], C. S. Fadley[7,8], A. X. Gray[7,8,9], J. Minar[10], J. Braun[10], H. Ebert[10], I. Di Marco[11], O. Eriksson[11], P. Thunström[12], G. H. Fecher[13], S. Ouardi[13], H. Stryhanyuk[13], E. Ikenaga[14], C.H. Back[2], G. van der Laan[15], and G. Panaccione[1*]

[1] *Istituto Officina dei Materiali (IOM)-CNR, Laboratorio TASC, in Area Science Park, S.S.14, Km 163.5, I-34149 Trieste, Italy*
[2] *Institut für Experimentelle Physik, Universität Regensburg, D-93040 Regensburg, Germany*
[3] *NIMS beamline Station at Spring-8, National Institute for Materials Science, Sayo, Hyogo 679-5148, Japan*
[4] *Japan Atomic Energy Agency, Sayo, Hyogo 679-5148, Japan*
[5] *Hiroshima Synchrotron Radiation Center, Hiroshima University, 2-313 Kagamiyama, Higashi-Hiroshima, Japan*
[6] *Advanced Electric Materials Center, National Institute for Materials Science, 1-1 Namiki, Tsukuba, Ibaraki 305-0044, Japan*
[7] *Department of Physics, University of California, Davis, California 95616, USA*
[8] *Material Sciences Division, Lawrence Berkeley National Laboratory, Berkeley, California 94720, USA*
[9] *Stanford Institute for Materials and Energy Sciences, SLAC National Accelerator Laboratory, Menlo Park, California 94025, USA*
[10] *Department of Chemistry, Ludwig Maximillian University, D-81377 Munich, Germany*
[11] *Department of Physics and Astronomy, Uppsala University, Box 516, SE-75120, Uppsala, Sweden*
[12] *Institute for Solid State Physics, Vienna University of Technology, 1040 Vienna, Austria*
[13] *Max Planck Institute for Chemical Physics of Solids, 01187 Dresden, Germany*
[14] *Japan Synchrotron Radiation Research Institute, SPring-8, Hyogo 679-5198, Japan*
[15] *Diamond Light Source, Chilton, Didcot, Oxfordshire OX11 0DE, United Kingdom*

## ABSTRACT

We report high-resolution hard x-ray photoemission spectroscopy results on (Ga,Mn)As films as a function of Mn doping. Supported by theoretical calculations we identify, over the entire 1% to 13% Mn doping range, the electronic character of the states near the top of the valence band. Magnetization and temperature dependent core-level photoemission spectra reveal how the delocalized character of the Mn states enables the bulk ferromagnetic properties of (Ga,Mn)As.

Understanding the electronic and magnetic properties of diluted magnetic semiconductors (DMSs) has been a major challenge in materials science over the last decade, and carrier-mediated ferromagnetism is undoubtedly one of the most discussed issues in DMS research [1-6]. The case of the (Ga,Mn)As valence band is prototypical: Whether the states near the Fermi level, $E_F$, are best described in terms of dispersive states fully merged with the GaAs valence band or if these states preserve the character of an impurity band, has been a central issue of intense activity in solid-state science, holding important ramifications for the development of future spintronics materials [3-6]. Both descriptions stand on various experimental and theoretical arguments that favor one or the other of these two visions [5-9], and more recently unified pictures, involving both ideas, have been presented [10-13]. Nonetheless, the crossover regime between impurity states at low doping and truly extended states at high doping, with possibly Bloch-like characteristics, i.e., the essence of how ferromagnetism is established in (Ga,Mn)As, is far from a unified and clear description. Such a description is intimately linked to a further central question: how important is localization vs. hybridization of carriers in the vicinity of $E_F$ as a function of doping? Theoretical modeling has so far only offered an inconclusive contribution to this issue, since disorder and strong electronic correlations render realistic electronic structure calculations difficult to perform. On the experimental front, full control of the system is hard to achieve and only a limited number of direct measurements of the valence band electronic structure of (GaMn)As have been reported [14,15]. The need to overcome the solubility limit to obtain stable conditions for ferromagnetic order, thus imposing off-equilibrium growth conditions, strongly influences the homogeneity of (Ga,Mn)As. Reliable experimental evidence exists, e.g., for a depletion zone near the interface in device-like structures that are responsible for dramatic changes in the hole-density profile, accompanied by a strong reduction of the Curie temperature, $T_C$ [16]. Moreover, the presence of both substitutional (acceptors) and interstitial (double-donor sites) impurities at high Mn concentrations, imposing their opposite contributions to the magnetic properties, leads to a profound difference between surface and bulk electronic properties [7,16].

In the present work, we exploit bulk-sensitive hard x-ray photoemission spectroscopy (HAXPES), corresponding to a severely suppressed (<4%) surface contribution [17,18], where the combined measure of valence band and core level, including temperature dependent core-level magnetic circular dichroism (MCD), allows a direct determination of the bulk density of states near $E_F$ as a function of Mn doping. Supported by theoretical calculations, we observe that, although the GaAs valence band is substantially modified by Mn doping: (*i*) a clear DOS feature with Mn 3*d* character is present, starting at a Mn doping level as low as 1%, (*ii*) the Fermi energy is located well above the GaAs host valence band top, and (*iii*) the delocalized character of the Mn states has a direct link with the (Ga,Mn)As ferromagnetic properties.

We have studied three ferromagnetic (Ga,Mn)As thin films (Mn 1% , 5%, and 13% doping, 18 to 50 nm of thickness). Samples were grown by molecular beam epitaxy on semi-insulating (001) oriented GaAs substrates, under low-temperature growth conditions. HAXPES measurements were carried out at SPring-8 in Hyogo, Japan using linearly horizontal polarized x-rays on undulator beamline BL15XU, and circularly polarized x-rays on undulator beamline BL47XU. The overall energy resolution (monochromator + analyzer) was set to 250 meV for core level and 150 meV for valence band experiments, as verified by measuring the spectra of the Au valence band at the Fermi level from a polycrystalline sample in thermal and electrical contact with the (Ga,Mn)As sample. The x-ray incidence angle as measured from the sample surface was fixed at 2°. Further details concerning the photoemission measurements, sample synthesis procedure, and calculation methods can be found in Ref. 19.

In Fig. 1 we show successive images of the valence-band region of pure GaAs and Mn-doped samples, with increasing magnification of the near $E_F$, where the experimental results [$h\nu$ = 5953 eV, Fig. 1(a-c)] are compared with the theoretical curves [Fig. 1(d-f)] based on a one-step model of angle-integrated photoemission. For the sake of comparison, the GaAs spectra have been aligned to the (GaMn)As spectra using the binding energy (BE) position of the As 4$s$ core level. The electronic structure calculations used for the one-step photoemission intensities are based on the fully-self-consistent combination of density-functional theory (DFT), the coherent-potential approximation (CPA) to describe the presence of the Mn dopant, and dynamical-mean-field theory (DMFT) as implemented within the multiple scattering Green's function formalism [12,20-24]. The calculations agree quantitatively with the experimental data over a wide BE range. The As 4$s$ states are located at ~10 eV BE and their effective hybridization with the Mn 3$d$ states is small as they are well separated in energy [Fig. 1(a, d)]. Closer to $E_F$, the Mn 3$d$ states of $e_g$ symmetry hybridize weakly with a mixture of As 4$p$ (mainly) and Ga 4$p$ states. Instead, the Mn 3$d$ states of $t_{2g}$ symmetry exhibit a more significant hybridization just below $E_F$.

In Fig. 1(c), high-resolution spectra measured in the vicinity of $E_F$ are displayed for pure GaAs, 1% and 13% Mn doping. Also shown are the difference spectra, with the pure GaAs spectrum is subtracted from the 13% and 1% Mn spectra (orange and grey, respectively), in order to highlight the explicit Mn contribution. In Fig. 1(f), the calculated difference spectra, GaAs subtracted from (GaMn)As, are presented for the LDA and DMFT calculations (violet and green curves, respectively). Both the experiment and calculation show a pronounced maximum caused by Mn *d*-states, arising from a mixture of mainly As 4*p* states, localized around the impurity, and Mn 4*p* states. The strong hybridization redistributes the spectral weight of the Mn 3*d* states over a wide energy range all the way up to $E_F$. The evolution as a function of doping shows an increasing intensity of the Mn band. Within the limit of our resolution, no shift of the Mn band is observed. The Mn *d* spectral weight is peaked at ~300 meV below $E_F$, in agreement with previous surface sensitive photoemission data [14], and recent hard x-ray angle-resolved photoemission [11]. It is important to emphasize that the Mn-related DOS near $E_F$, although significantly smaller, is nonzero within the energy resolution. The diluted nature of Mn in the host GaAs matrix is confirmed by the spectroscopic fingerprint of the difference spectra in Fig. 1(c): the Mn states in (Ga,Mn)As retain their Gaussian lineshape (broadened when the doping value increases) up to a Mn doping as high as 13%, a value safely considered to display 'metallic' behavior, whereas for MnAs, a truly metallic system, one obtains a clear Fermi edge [19]. The Mn band shows neither a crossover towards a truly metallic Fermi level, nor are clustering-related features found in core-level photoemission spectroscopy (PES) and x-ray absorption spectroscopy [19].

Having ascertained the nature and evolution of the Mn states in the vicinity of $E_F$, we further address their role in the ferromagnetic properties of the (Ga,Mn)As system by means of core-level studies. In previous work, core-level PES in the hard x-ray regime was able to demonstrate, via comparison with Anderson impurity model (AIM) model calculations, the importance of hybridization vs. localization of *d*-electrons in (Ga,Mn)As [25,26]; to such measurements, we have now added magnetic sensitivity by using circularly polarized x-rays, thus enabling MCD-PES. Figure 2 shows the core-level HAXPES ($hv$ = 7940 eV) for a 13% Mn doped GaAs sample in the ferromagnetic state (below its Curie temperature, $T_C$ = 80 K), and compared to model calculations. Figure 2(a) shows a survey spectrum that locates the Mn 2$p$ doublet of interest, and identifies the well- and poorly-screened features and multiplet structure of the 2$p_{3/2}$ and 2$p_{1/2}$ core levels, in agreement with previous results [25,26]. Figure 2(b) shows the MCD for the full Mn 2$p$ region, with a very large magnetic asymmetry of ~13%. It has previously been noted that the line shape of the Mn 2$p$ core level does not change while varying the Mn concentration from 1% to 13%, which is strong evidence against an important contribution of interstitial Mn atoms in our bulk-sensitive spectra [25]. The calculations in Fig. 2(c), using the AIM model [19], shows that the 2$p_{3/2}$ structure contains a well-screened peak at the high kinetic energy (low BE) side and a poorly-screened peak at 1-2 eV lower kinetic energy (higher BE), with opposite dichroism [25,26]. The well-screened peak is found to have mainly 2$p^5$3$d^6h^2$ character, while the poorly screened peak has mainly 2$p^5$3$d^5h$ character, where $h$ denotes a hole state. In the ground state, the 3$d^5h$ state is pinned to $E_F$, and it is the 3$d^6h^2$ state which is pulled down in energy in the final state due to the 2$p$ core-hole potential [25]. In the calculation, the local ground state properties are primarily reflected by two parameters: The transfer integral, $V$, responsible for the hybridization (mixing) and the on-site 3$d$ electron-repulsion energy, $U$, i.e., the Hubbard $U$. For small $U/V$ the electrons are delocalized, whereas for large $U/V$ the electrons are localized on the atomic sites. The experimental MCD of the HAXPES agrees very well with the calculated spectra for a hybridization parameter 2.0 ≤ V ≤ 2.5 eV. It is important to emphasize that the most intense magnetic signal, at the energy location corresponding to the leading well-screened peak, is linked to the most delocalized electronic *d*-character, an attribute already put forward by the results of Dobrowolska *et al*. [27] and in good agreement with Richardella *et al.* [9].

Additional evidence of the important role played by the electron hybridization of the Mn 3*d* states is obtained from temperature dependent HAXPES. Knowing that the carrier concentration in a heavily doped semiconductor, such as (GaMn)As, does not change significantly with temperature [27], a shift of the core levels and/or the valence band spectra vs. temperature is associated with a change in the screening efficiency of the delocalized electrons. The HAXPES results are shown in Fig. 3 for the Ga 2*p* and Mn 2*p* core levels (13% Mn doping), and compared to the valence band. While the Ga 2*p* peak shifts ~80 meV to higher BE between 20 K and room temperature, the measured valence-band shift is less than half of this value. This temperature dependence implies an increased Mn 3*d*-As 4*p* hybridization, and hence a more efficient hopping favoring ferromagnetism [26-29]. Moreover, the low BE feature of the Mn $2p_{3/2}$ core level (associated with the more delocalized $d^6h^2$ states) shifts with temperature by ~35 meV, whereas no shift of the broad component at higher BE (corresponding to the more localized $d^5h$ states) is detected. This observation confirms that the screening is associated exclusively to the more hybridized Mn 3*d* states, $d^6h^2$, while the localized Mn $d^5h$ states, in the vicinity of $E_F$, do not participate in the screening process. A further confirmation of this picture comes from the calculated MCD-PES in Fig. 2(b), where the energy separation between the sharp negative peak and broad positive peak in the Mn $2p_{3/2}$ increases with *V*; this again implies a more efficient screening due to delocalized electrons. Therefore, the observed BE shifts are not due to surface effects. It is important to underline that the HCl chemical etching performed prior to introducing the samples in vacuum passivates the surface states and results in a complete unpinning of the Fermi level [19].

The physical scenario emerging from our combined observations and calculations is presented in Fig. 4. The GaAs valence band is substantially modified by the introduction of Mn atoms, even for doping values as low as 1%. The largest modification of the GaAs host band is found over 2-4 eV BE. For all Mn doping levels studied, covering 1-13 % of the measured (Ga,Mn)As samples, the energy separation from the top of the GaAs-host states to $E_F$ is in the order of 50 meV, after taking into account the experimental energy resolution, convolution with the Fermi function, and possible recoil effects [19]. The evolution as a function of doping of the Mn-derived states shows an increasing intensity, but no shift, neither a crossover towards a truly metallic Fermi level. The Mn $d$ spectral weight is peaked at ~300 meV below $E_F$. We note that in a real impurity system, i.e., when disorder is included, Bloch states are never formed and the concept of delocalized states in the Mn-derived band with a clear $E$ vs. $k$ relation cannot be fully defined. Thus the term 'band' does not imply electronic states of well-defined crystal momentum, while the DOS remains a well-defined quantity. In the present case, angle-integrated HAXPES measures only the DOS: due to the high photon energy used ($hv \geq 5$ keV), band structure effects can be neglected [16].

Knowing that the key differences between the valence band and the impurity band model are found in: (*i*) the description of the nature of the electronic wave functions mediating magnetism [4] and (*ii*) the location of $E_F$ with respect to the GaAs host matrix, the present experimental observations of a persistent diluted character of Mn in GaAs, together with $E_F$ located well above the GaAs valence band top, as found by Ohya *et al*. [8], suggest that the existence of a Mn-derived band which is not detached from the GaAs bands, could be compatible at the same time with localization (impurity band) character.

In conclusion, our bulk sensitive results provide experimental and theoretical evidence of a persistent diluted character of Mn in GaAs, up to high concentration doping. We confirm the presence and importance of electronic correlations in the vicinity of the Fermi level, and at the same time the fundamental role of the delocalized Mn-derived electron states for the ferromagnetic character of the system.

**Acknowledgements**

Two of us (AXG and CSF) gratefully acknowledge the support of the US Department of Energy under Contract No. DE-AC02-05CH11231. Financial support from the German funding agencies DFG (FOR 1346, SFB 689, EB 154/23 and 18), the German ministry BMBF (05K10WMA) and the Austrian Science Fund (FWF project ID I597-N16) is also gratefully acknowledged (J.M., J.B., H.E. and P.T.). I.D.M. and O.E. are grateful to the Swedish Research Council (VR), Energimyndigheten (STEM), the KAW Foundation and ERC (247062 - ASD). We like to thank M. I. Katsnelson for useful discussions.

**FIGURE CAPTIONS**

**FIG. 1.** (Color online) (**a-c**) HAXPES spectra ($h\nu$ = 5953 eV) for different Mn doping in GaAs. For the sake of comparison, the GaAs spectra have been aligned to the (GaMn)As spectra using the BE position of the As 4s core level. No background subtraction was applied. **(a)** Extended valence band PES ($T = 20$ K), including the As 4s, Ga 4s, and As 4p shallow core levels of GaAs(100) and 13% Mn doped GaAs. **(b)** Zoom of the valence band region ($T = 100$ K), showing the spectra from pure GaAs (black dots), 1% and 13% Mn doped GaAs (red and blue dots, respectively). **(c)** High-resolution spectra measured in the vicinity of $E_F$. Difference spectra, corresponding to the Mn contribution only, are shown in orange [13% Mn spectrum (blue dots) minus pure GaAs spectrum (black dots)] and grey [1% Mn spectrum (red dots) minus pure GaAs spectrum (black dots)]. **(d)** Calculated angle-integrated PES (including matrix elements) for photon energy and geometry (p-polarization) as used in the experiment. **(e)** Calculated valence-band spectra of GaAs(100) using LDA (black curve) and (Ga,Mn)As (13%) using both LDA (red curve) and DMFT (blue curve). **(f)** Zoom of the vicinity of $E_F$ with calculated difference spectra, as in panel **(c)**, for LDA (violet curve) and DMFT (green curve).

**FIG. 2.** (Color online) **(a)** HAXPES wide energy scan ($h\nu = 7940$ eV, linear polarization, $T = 20$ K) of 13% Mn doped GaAs. Main lines positions are indicated. The inset shows a zoom of the Mn 2p core level with the $2p_{1/2}$ and $2p_{3/2}$ multiplet structure, where the well-screened and poorly-screened peaks are indicated. For an explicit definition of well and poor screened peaks in (GaMn)As see ref [25]. **(b)** Magnetic circular dichroism (MCD) in HAXPES. Polarization dependent Mn 2p spectra (blue and red dots for right- and left-circular polarization, respectively), measured at $T = 20$ K, i.e., in the ferromagnetic state below $T_C$; the dashed thin line represents the background. In the lower panel the difference spectrum (MCD), representing the magnetic signal (open white circles) is shown; the green line is a smoothed curve. The value of the maximum magnetic asymmetry, defined as $(I^+ - I^-)/(I^+ + I^-)$, where $I^+$ ($I^-$) is the intensity measured with Circ+ (Circ−), is indicated. **(c)** Calculations for an Anderson impurity model with varying hybridization $V$, as well as a fully localized $d^5$ state, and an itinerant electron model. The calculated curves are offset for clarity. The best agreement is for a hybridization parameter $2.0 \leq V \leq 2.5$ eV.

**FIG. 3.** (Color online) Temperature dependence of the (Ga,Mn)As (13% Mn) PES. (**a**) Relative BE shift of the GaAs host valence band, Ga 2$p$ core level, Mn 2$p_{3/2}$ low energy feature, and Mn 2$p_{3/2}$ broad peak; the BEs at 20 K are used as reference. (**b**) Mn 2$p_{3/2}$ core level measured at 20 K and at RT.

**FIG. 4.** (Color online) Schematics summarizing our conclusions regarding the electronic structure of (Ga,Mn)As. (**a**) Experimental valence band (smoothed) at $h\nu = 5953$ eV. GaAs host band (blue curve) and (Ga,Mn)As band with 13% Mn (red curve). The difference between both spectra identifies the Mn-derived states (dark green). The dashed vertical bar marks the position of $E_F$; the center of the Mn band is located at 300 meV BE. (**b**) The bottom panel shows a zoom of the region near $E_F$. The GaAs host band does not cross $E_F$ (energy separation of 50 meV) but the Mn-induced states have a finite non-zero density at $E_F$. The color shading illustrates the difference in electronic character of the Mn $d$-states as derived from our theoretical calculations: near $E_F$ a $3d^5h$ localized character is found, whereas $3d^6h^2$ delocalized character is assigned to the dark green region. The latter corresponds to the hybridization responsible for mediating the ferromagnetism.

**Fig. 1**

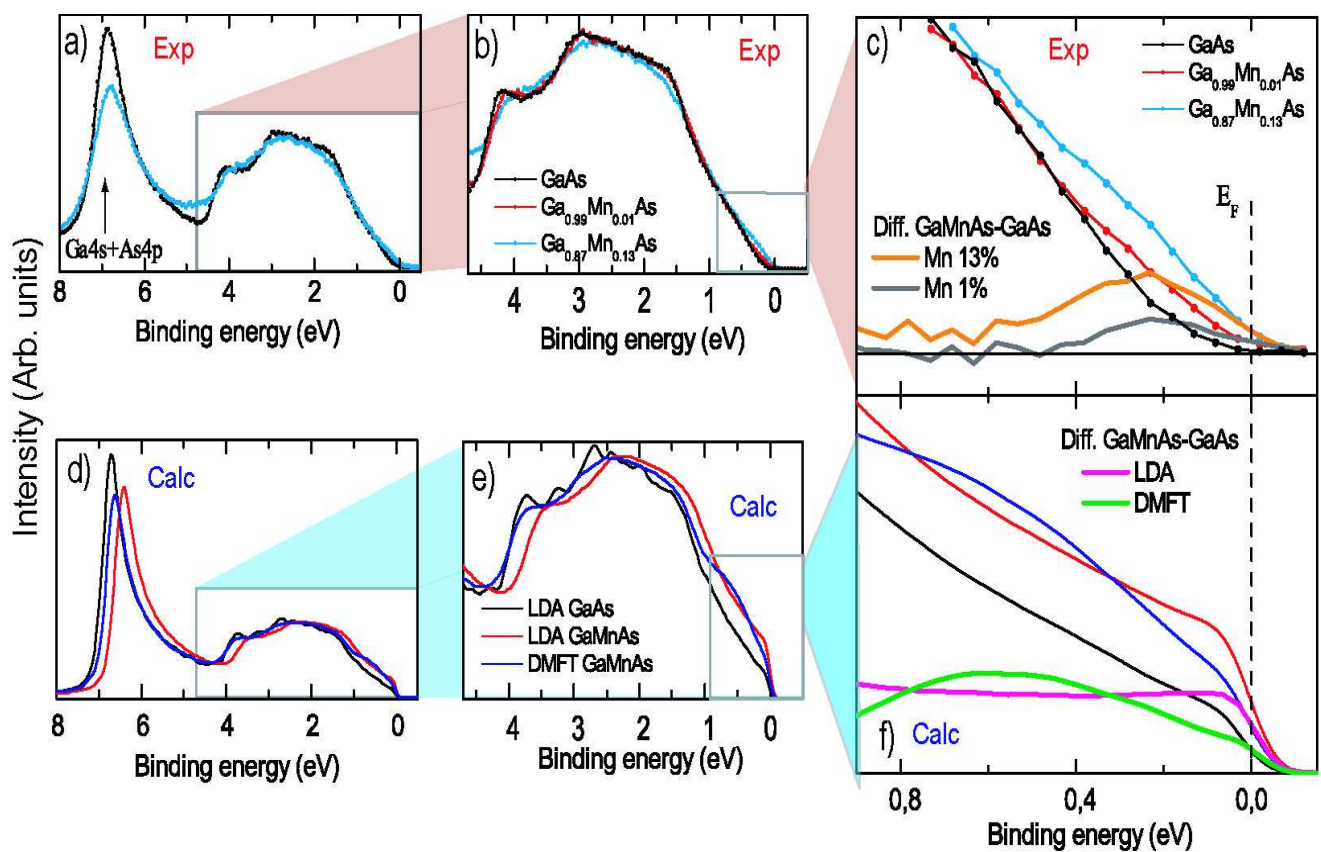

**Fig. 2**

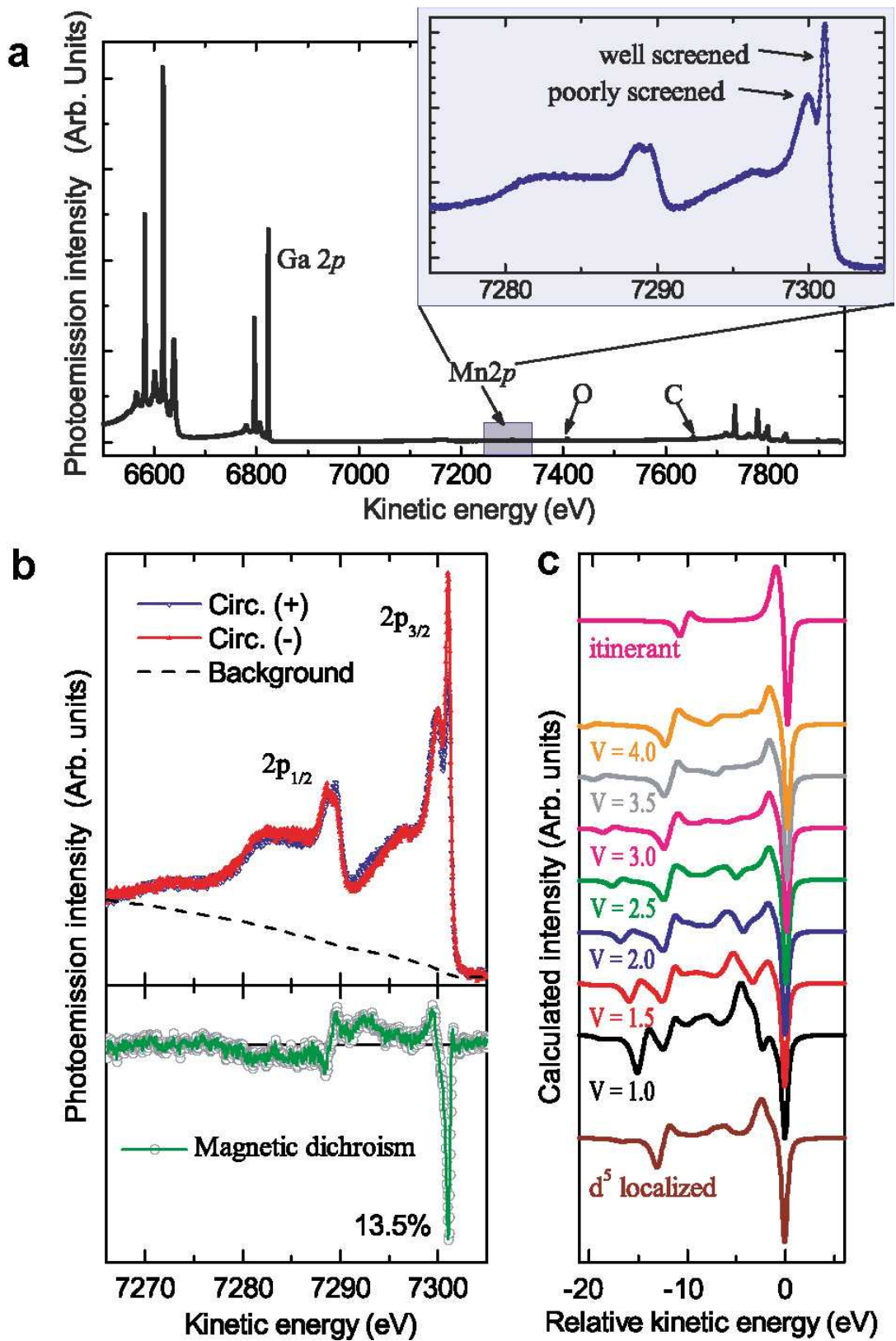

**Fig. 3**

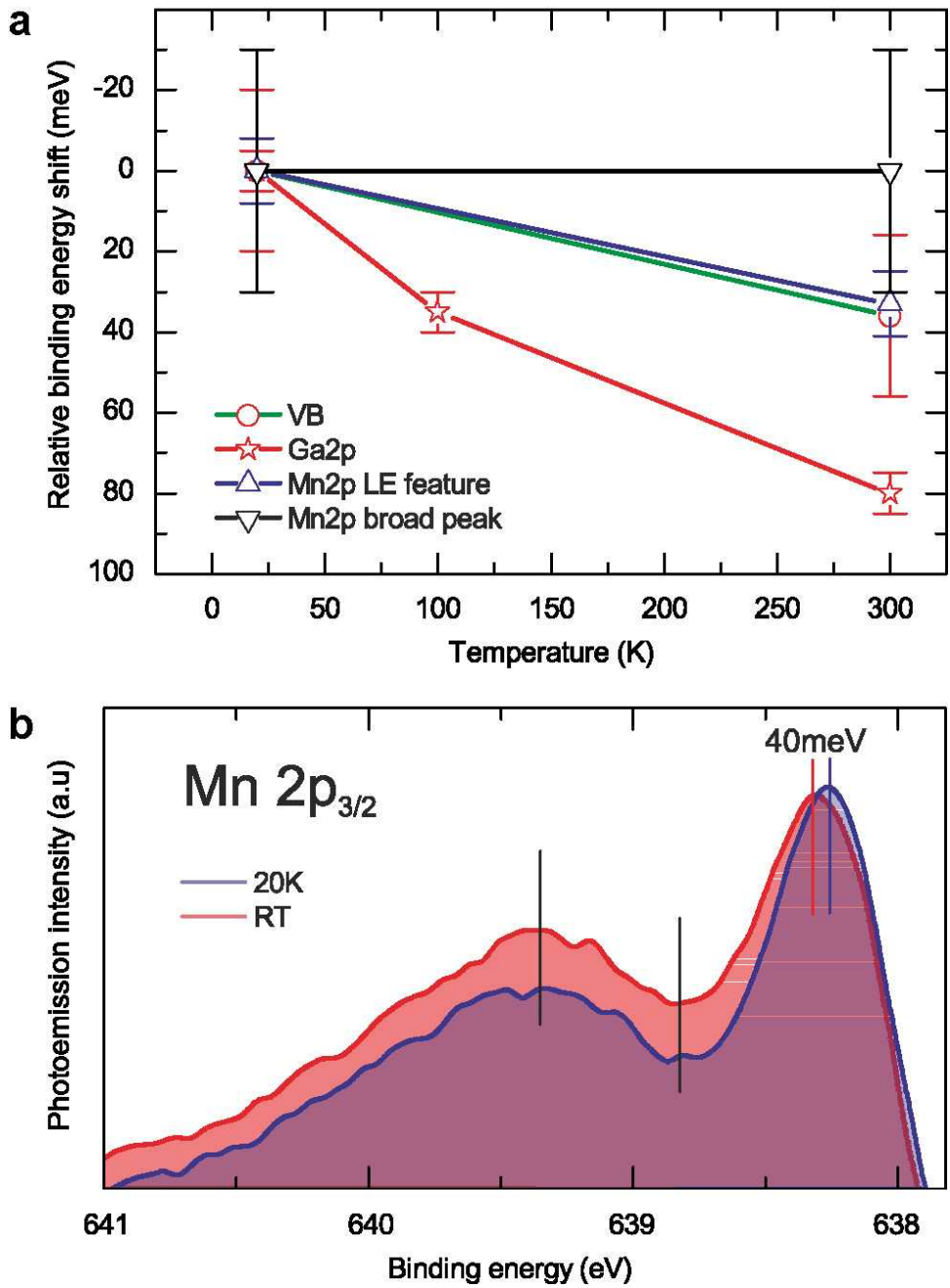

**Fig. 4**

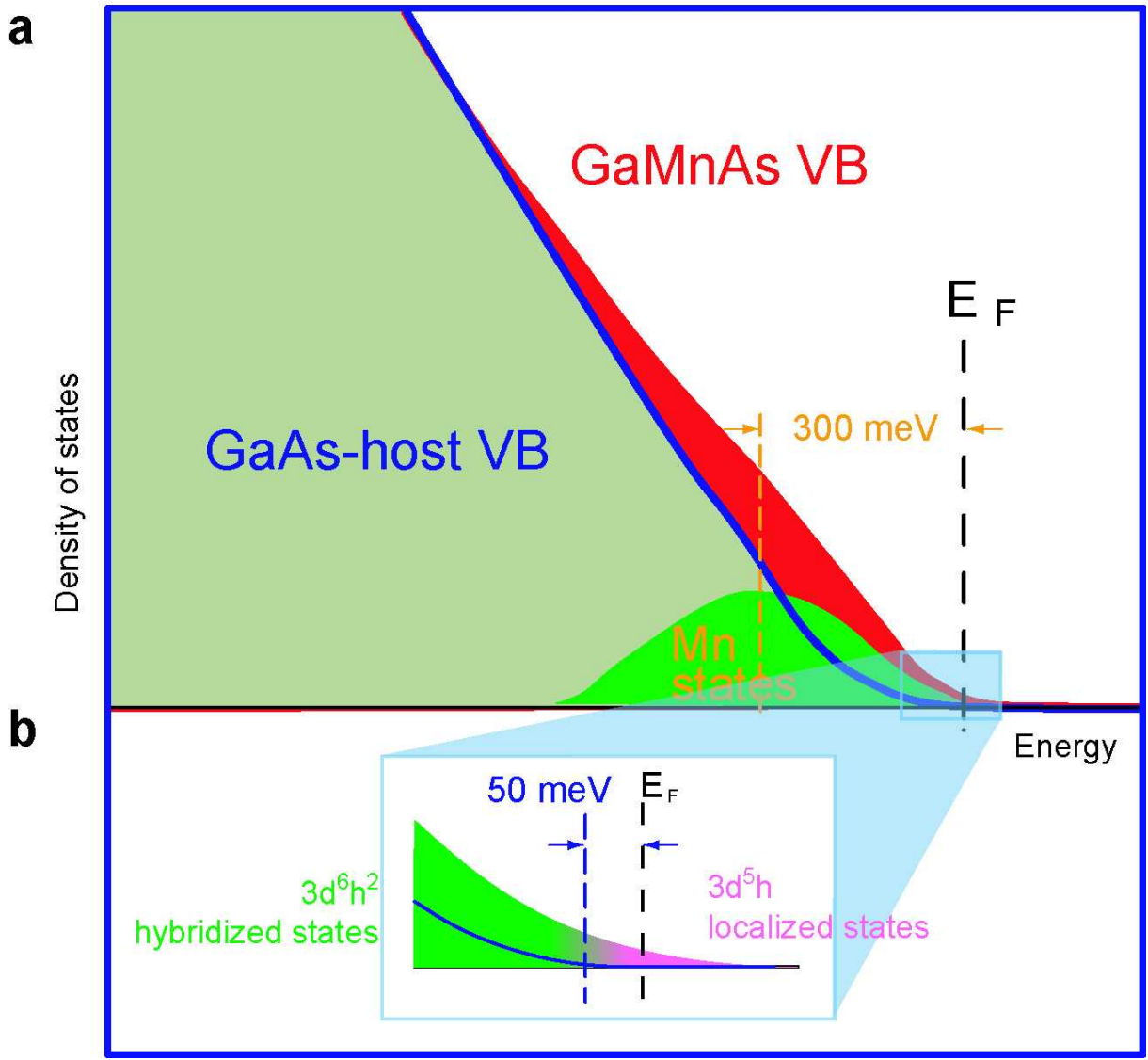